\documentclass[sigconf,nonacm,table,prologue]{acmart}
\AtBeginDocument{%
  }

\setcopyright{acmlicensed}
\copyrightyear{2026}
\acmYear{2026}
\acmDOI{XXXXXXX.XXXXXXX}
\acmConference[Sensys '26]{Proceedings of the ACM on Embedded Artificial Intelligence and Sensing Systems}{May 11--14,
  2026}{Saint-Malo, France}
\acmISBN{xxxx}

\PassOptionsToPackage{table,prologue}{xcolor}    
\usepackage{xcolor}                              
\usepackage{graphicx,amsmath,enumitem,tabularx,makecell,diagbox,longtable}
\usepackage{multirow,hhline,multicol}
\usepackage{caption}                             
\usepackage{subcaption}
\usepackage{array}
\usepackage{listings}                            
\usepackage[export]{adjustbox}                   
\usepackage{todonotes,tcolorbox,pifont,bbding}
\usepackage{tikz}

\newcommand{\etc}{\textit{etc}}
\newcommand{\sys}{\texttt{AutoBridge}}

\definecolor{codegreen}{rgb}{0,0.6,0}
\definecolor{codegray}{rgb}{0.5,0.5,0.5}
\definecolor{codepurple}{rgb}{0.58,0,0.82}
\definecolor{backcolour}{rgb}{0.95,0.95,0.92}
\lstdefinestyle{mystyle}{
    backgroundcolor=\color{backcolour},   
    commentstyle=\color{codegreen},
    keywordstyle=\color{magenta},
    numberstyle=\tiny\color{codegray},
    stringstyle=\color{codepurple},
    basicstyle=\ttfamily\footnotesize,
    breakatwhitespace=false,         
    breaklines=true,                 
    captionpos=b,                    
    keepspaces=true,                 
    numbers=left,                    
    numbersep=5pt,                  
    showspaces=false,                
    showstringspaces=false,
    showtabs=false,                  
    tabsize=2
}
\colorlet{punct}{red!60!black}
\definecolor{background}{HTML}{EEEEEE}
\definecolor{delim}{RGB}{20,105,176}
\colorlet{numb}{magenta!60!black}
\lstdefinelanguage{json}{
    basicstyle=\ttfamily\footnotesize,
    numbers=left,
    numberstyle=\tiny\color{codegray},
    stepnumber=1,
    numbersep=8pt,
    showstringspaces=false,
    breaklines=true,
    frame=lines,
    backgroundcolor=\color{background},
    literate=
     *{:}{{{\color{punct}{:}}}}{1}
      {,}{{{\color{punct}{,}}}}{1}
      {\{}{{{\color{delim}{\{}}}}{1}
      {\}}{{{\color{delim}{\}}}}}{1}
      {[}{{{\color{delim}{[}}}}{1}
      {]}{{{\color{delim}{]}}}}{1},
}
\begin{document}

\title[AutoBridge]{AutoBridge: Automating Smart Device Integration with Centralized Platform}

\author{Siyuan Liu}
\email{sliu268@connect.hkust-gz.edu.cn}
\affiliation{%
  \institution{Hong Kong University of Science and Technology (Guangzhou)}
  \city{Guangzhou}
  \state{Guangdong}
  \country{China}
}

\author{Zhice Yang}
\email{yangzhc@shanghaitech.edu.cn}
\affiliation{%
  \institution{ShanghaiTech University}
  \city{Shanghai}
  \country{China}
}

\author{Huangxun Chen}
\email{huangxunchen@hkust-gz.edu.cn}
\affiliation{%
  \institution{Hong Kong University of Science and Technology (Guangzhou)}
  \city{Guangzhou}
  \state{Guangdong}
  \country{China}
}
\thanks{Corresponding Author: Huangxun Chen.}

\renewcommand{\shortauthors}{Liu et al.}

\begin{abstract}

Multimodal IoT systems coordinate diverse IoT devices to deliver human-centered services.
The ability to incorporate new IoT devices under the management of a centralized platform is an essential requirement.
However, it requires significant human expertise and effort to program the complex IoT integration code that enables the platform to understand and control the device’s functions.
Therefore, we propose \sys~to automate IoT integration code generation.
Specifically, \sys~adopts a divide-and-conquer strategy: it first generates device control logic by progressively retrieving device-specific knowledge, then synthesizes platform-compliant integration code using platform-specific knowledge.
To ensure correctness, \sys~features a multi-stage debugging pipeline, including an automated debugger for virtual IoT device testing and an interactive hardware-in-the-loop debugger that requires only binary user feedback (“yes/no”) for real-device verification. We evaluate \sys~on a benchmark of 34 IoT devices across two open-source IoT platforms. 
The results demonstrate that \sys~can achieves an average success rate of 93.87\% and an average function coverage of 94.87\%, without any human involvement. With minimal binary "yes/no" feedback from users, the code is then revised to reach 100\% function coverage.
A user study with 15 participants further shows that \sys~outperforms expert programmers by 50\%–80\% in code accuracy, even when the programmers are allowed to use commercial code LLMs.

\end{abstract}

\ccsdesc[500]{Human-centered computing~Ubiquitous and mobile computing systems and tools}
\ccsdesc[500]{Human-centered computing~Natural language interfaces}


\keywords{IoT Management, Large Language Model, IoT Program Synthesis}

\maketitle

\section{Introduction}

\begin{figure}
    \centering
    \includegraphics[width=1\linewidth]{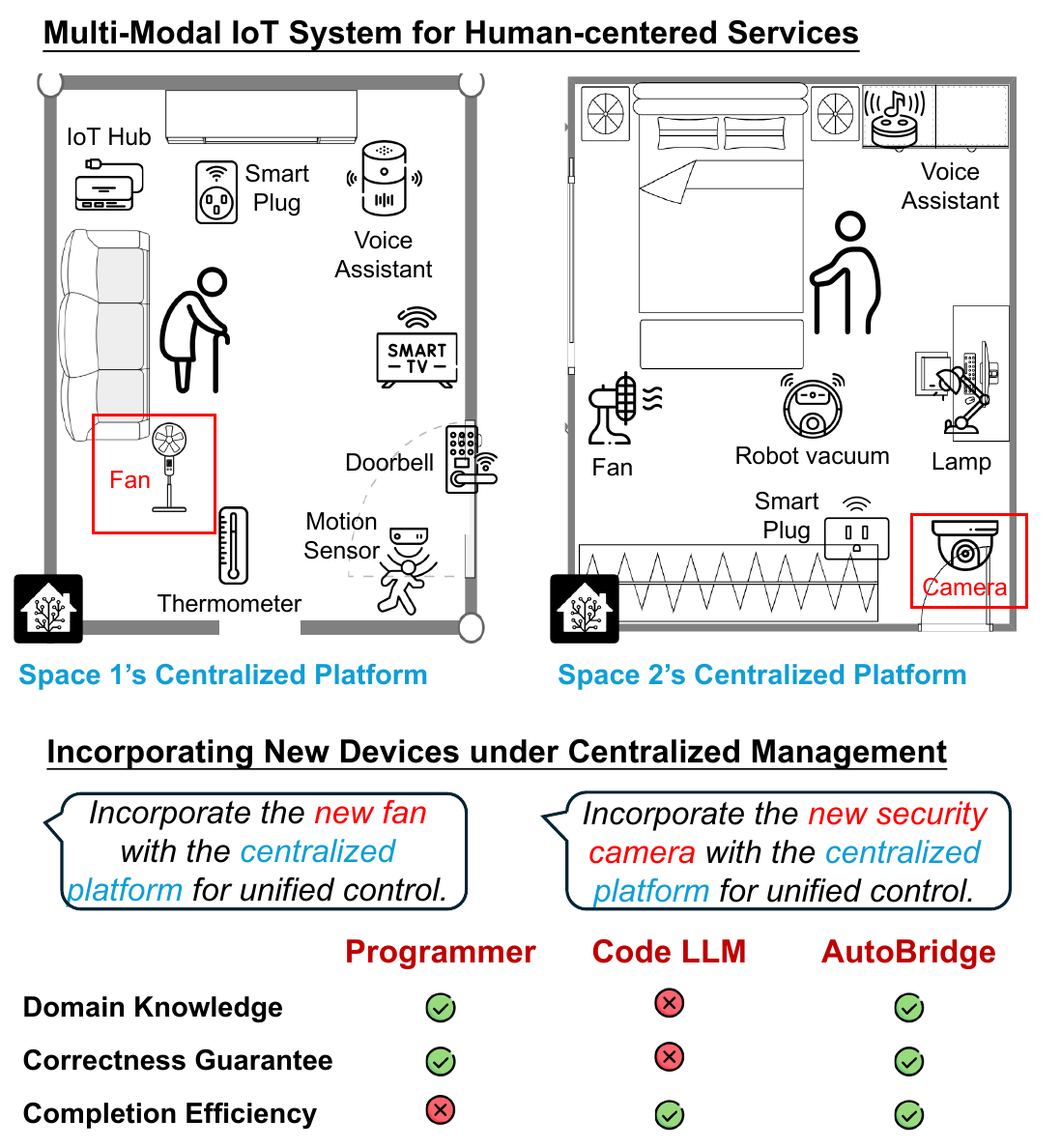}
    \caption{Personalized multi-modal IoT systems need to support dynamic upgrades and the addition of new IoT devices, which requires programming integration code to incorporate these devices with the centralized platform. 
    }
    \Description{The main usage scenario of the proposed system.}
    \label{fig:intro-fig}
\end{figure}


Over the past few decades, IoTs have been widely integrated into everyday life.
As illustrated in Figure~\ref{fig:intro-fig}, to serve human-centered needs, the IoT system encompasses multi-modal devices distributed throughout the space, with a centralized platform, e.g.~\cite{homeassistant2024,openHAB,ifttt} integrating all device control interfaces for unified coordination. 
However, no single device combination can satisfy all needs. Each user has unique requirements and budget constraints.
These needs often evolve over time, leading to the addition or upgrade of devices within the space. 
A practical hurdle in incorporating a newly introduced device into the existing platform is the need to implement IoT integration code—custom code that is plugged into the centralized platform to help establish a connection with the device, retrieve data from it, and send control commands to it.

In the past, we typically relied on programmers to implement integration code. However, this has proven to be a daunting task, as it requires gathering technical details from device documentation and understanding the platform’s technical framework—particularly the onboarding procedures for new IoT integrations~\cite{ha2024integration}.
Even for experienced developers, the process demands significant effort. 
In our user study, more than half of the expert programmers can not write correct code within 40 minutes, even for a smart light bulb with relatively simple functionality.
This is why no single centralized platform includes integration code for all smart devices.
For example, Home Assistant's developer provides integration code for certain Yeelight LED bulbs, such as the Yeelight LED Bulb White and Yeelight LED Bulb Color E26, but these codes are not fully compatible with the Yeelight LED Bulb 1S (Color) due to functional differences, e.g., a new rhythmic mode.
The workload required to manually support the vast number of existing and emerging IoT devices is simply unmanageable.



\begin{figure}
    \centering
    \includegraphics[width=1\linewidth]{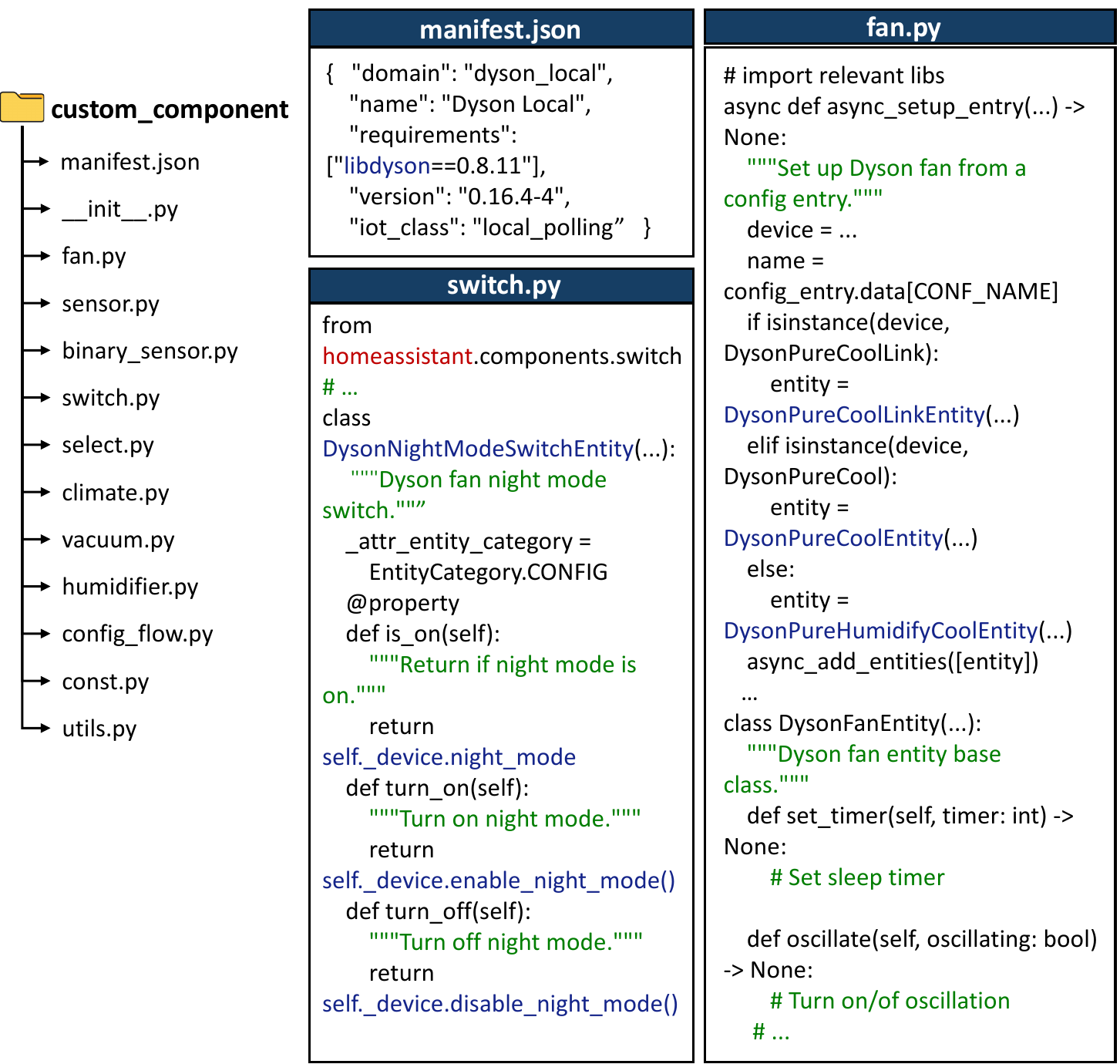}
    \caption{\textbf{Integration code example} showcases its "bridge" nature between the platform (i.e., Home Assistant) and the device (i.e., Dyson Fan). The \textbf{left figure} illustrates the structure of a complete integration. The \textbf{middle/right figures} display some code snippets. 
    }
    \label{fig:code-exp}
    \Description{Integration Code Example}
    \vspace*{-10pt} 
\end{figure}

The emergence of code LLMs~\cite{nijkampcodegen,li2022competition,li2023starcoder,guo2024deepseek,roziere2023code,team2024codegemma,luowizardcoder,dou2024stepcoder} presents a promising solution to this challenge. If IoT integration code between a platform and a device can be automatically generated, it would enable more seamless and flexible expansion of existing multi-modal IoT systems in Figure~\ref{fig:intro-fig}. We can incorporate more IoT devices to enhance sensing and human-centered services without the need to wait for time-consuming manual development of integration code. However, directly applying code LLMs does not work well for synthesizing workable IoT integration code, as it entails three key technical challenges. 
\begin{itemize}[leftmargin=*]
    \item \textbf{CH1: Complexity of IoT integration code}. Unlike existing code generation benchmarks~\cite{austin2021program, chen2021evaluating}, which typically involve well-defined modules or functions with 5-40 lines of code (LOC), IoT integration code often exhibits a much more complex structure (see Figure~\ref{fig:code-exp}) and spans 300-3000 LOC (see Table~\ref{tab:compare_code_loc}), making it hard to generate it with high accuracy. 
    \item \textbf{CH2: Heavy Domain Knowledge for IoT Integration}. IoT integration programming demands extensive domain expertise. At its core, integration code serves as a bridge. On one hand, it must conform to the platform’s framework specifications. On the other hand, it must accurately identify and utilize the device’s interfaces to support intended operations (see Figure~\ref{fig:code-exp}). As IoT devices become more feature-rich, programming correct integration code requires even deeper domain knowledge. Blindly feeding code LLMs with domain knowledge through RAG~\cite{gao2023retrieval,fan2024survey} offers limited performance improvement, as the overwhelming volume of information generally makes it more difficult for the LLM to reason about where and how to apply the right knowledge precisely—even with an extended context window. 
    \item \textbf{CH3: Costly Validation and Debugging of IoT Integration.} Unlike prior coding tasks~\cite{austin2021program, chen2021evaluating,shen2025autoiot}, integration code cannot be fully verified in a simple execution sandbox environment (e.g., an IDE or code interpreter~\cite{openaicodeinterpreter}), because its correctness fundamentally depends on real-time interaction with external hardware, i.e., whether the deployed code enables the platform to fully manage the device and invoke all intended functions. 
    However, involving actual devices and human feedback in the debugging loop is often costly and inefficient. Thus, a comprehensive solution is needed to detect and correct errors in the generated integration code at minimal cost.
\end{itemize}

To address the above challenges, we propose \sys, a user-friendly programming agent that automates the integration of smart IoT devices with centralized platforms.
To tackle \textbf{CH1\&CH2}, \sys~adopts a divide-and-conquer strategy for code generation. Specifically, the code agent first incorporates device-specific knowledge to satisfy IoT devices' functional control requirements, and then refines the code using platform-specific knowledge to ensure compatibility with the target platform. Compared to attempting to absorb both device and platform knowledge simultaneously and generate the code in a single step, this strategy significantly improves generation accuracy.
To tackle \textbf{CH3}, we design an efficient multi-stage debugging framework. Specifically, we introduce an automated debugger based on virtual devices to iteratively refine and validate the code at the platform level without involving real hardware. Once the code passes virtual validation, we transition to a hardware-in-the-loop debugging phase, where a human inspector ensures the code behaves correctly on the actual IoT device and provides simple binary feedback ("yes" or "no"). This staged design strikes a balance between automation and accuracy, significantly reducing development costs.
An overview of \sys~is shown in Figure~\ref{fig:overview}, which consists of three sub-agents: an integration code generator, an automated debugger, and a hardware-in-the-loop debugger.

We implement \sys~and conduct two studies to evaluate its performance: 
(1) We curate a comprehensive integration code benchmark, covering 34 IoT devices across 2 popular centralized platforms using different programming languages (Python and Java). This benchmark is open-sourced to support future research\footnote{https://anonymous.4open.science/r/AutoBridge} (\S\ref{sec:benchmarkeval}). 
(2) We conduct a user study involving 15 participants, including both non-programmers and programmers, to evaluate the system's usability in integrating two representative devices into a chosen platform (\S\ref{sec:userstudy}).
The results demonstrate that \sys~can achieves an average success rate of 93.87\% and an average function coverage of 94.87\%, without any human involvement. With minimal binary "yes/no" feedback from users, the code is then revised to reach 100\% function coverage. 
The user study confirms AutoBridge's usability and efficiency.
It outperforms programmers’ manual coding performance by 50\%–80\% in terms of code accuracy, even when programmers are allowed to use commercial code LLMs.
It also enables non-programmers to generate the desired integration code. 
Moreover, subjective feedback from participants provides valuable insights into the future development of IoT code agents. 

Our contributions can be summarized as follows:
\begin{itemize}[leftmargin=*]
    \item To the best of our knowledge, this work is the first to fully automate the generation of IoT integration code—spanning hundreds of lines and featuring complex structures—with high accuracy and strong correctness guarantees. 
    \item In our proposed design, \sys, we equip the code generator with access to relevant knowledge and empower it with autonomous decision-making and divide-and-conquer strategy to incorporate domain knowledge, ensure generalization across IoT devices and platforms of varying complexity. In addition, we design a systematic scheme to ensure the correctness of the generated integration code in an efficient manner.
    \item We implement \sys~and conduct extensive evaluations. The results demonstrate \sys's high effectiveness, efficiency, usability in generating IoT integration code, outperforming the baselines. 
\end{itemize}

\section{Background and Motivation}

\begin{table}
    \small
    \renewcommand{\arraystretch}{1.2}
    \centering
    \begin{tabular}{|c|c|c|}
        \hline
        \textbf{Type} & \makecell{\textbf{Code Generation}\\\textbf{Benchmark}} & \makecell{\textbf{Average \#LOC}\\ \textbf{per Task}} \\
        \hline
        \multirow{7}{*}{\textbf{General Code}} & MBPP~\cite{austin2021program} & 6.68 \\
        \cline{2-3}
        & HumanEval~\cite{chen2021evaluating} & 8.95 \\
        \cline{2-3}
        & APPS~\cite{hendrycks2measuring} & 16.46 \\
        \cline{2-3}
        & DS-1000~\cite{lai2023ds} & 3.66 \\
        \cline{2-3}
        & \makecell{ComplexCodeEval\\-Java/Python~\cite{feng2024complexcodeeval}} & 31.72 / 38.21 \\
        \hline
        \multirow{3}{*}{\textbf{Integration Code}} & \makecell{openHAB\\-Tier1/Tier2/Tier3 (Java)} & 391 / 723 / 672.57\\
        \cline{2-3}
        & \makecell{HomeAssistant\\-Tier1/Tier2/Tier3 (Python)} & 709 / 1330 / 2882 \\
        \hline
    \end{tabular}
    \caption{Comparison between general code and integration code. The latter includes official integration codes from centralized platforms\cite{homeassistant2024,openHAB}. Here, Tier 1 denotes devices with 1–6 functions, Tier 2 denotes 7–10 functions, and Tier 3 denotes 11+ functions. LOC stands for Lines of Code. 
    }
    \label{tab:compare_code_loc}
    \vspace*{-26pt} 
\end{table}

We first revisit prior efforts on code LLMs and LLM-based code agents in IoT and embedded-related programming. Then, we present the results of our preliminary study, highlighting the challenges in accurately and automatically generating IoT integration code. 

\subsection{Code LLM and Domain-specific Code Agents}

Recently, we have witnessed the rapid development of large language models (LLMs)\cite{ouyang2022training, claude3, touvron2023llama, PaLM, dong2024survey, wei2022chain, yao2022react, yao2024tree}.
One particularly compelling application of LLMs is code generation\cite{jiang2024survey, he2025llmagentsurvey, liang2024AItoolsurvey}, which generally falls into two main paradigms:
(1) Pretraining specialized code LLMs from scratch~\cite{guo2024deepseek, li2023starcoder, li2022competition, nijkampcodegen}; 
(2) Fine-tuning existing LLMs on code-specific corpus~\cite{roziere2023code, luowizardcoder, dou2024stepcoder}.
Despite extensive efforts, most works are evaluated on general-purpose coding benchmarks~\cite{austin2021program, chen2021evaluating, hendrycks2measuring, lai2023ds, feng2024complexcodeeval}, typically spanning only 3-40 LOC.
Some recent studies~\cite{zhang-etal-2023-repocoder, phan2024repohyper} have started exploring repository-level code tasks, but their focus remains primarily on code completion rather than full code generation.
As a result, these approaches are not readily applicable to generating IoT integration code.

Many recent studies~\cite{hongmetagpt,shen2025autoiot,englhardt24exploring,ChatIoT,sasha2024,shen2025gpiot,gong2024poster,liu2025tasksense} have made efforts to develop LLM-based code agents that coordinate LLMs with other associated techniques to better meet the requirements of domain-specific code generation. 
Knowledge augmentation is a common practice, as domain-specific knowledge is often sparse in the intrinsic knowledge of code LLMs.
Prior works~\cite{shen2025autoiot,ChatIoT,sasha2024} incorporate RAG technique~\cite{gao2023retrieval,fan2024survey} to help generate accurate sensor-based HAR programs or IoT automation scripts.
Our work draws inspiration from these efforts but also recognizes that IoT integration code exhibits a significantly different nature. First, it typically involves longer code with more complex structures, make it harder for generation. Second, it requires more intricate and domain knowledge, making simple RAG is insufficient to support accurate generation. Third, contrary to data processing code~\cite{shen2025autoiot}, IoT integration code is designed to interact with real hardware. As a result, self-examination by LLMs~\cite{tian2024debugbench} or only debugging in code interpreters~\cite{shen2025autoiot} is insufficient for ensuring its correctness. However, hardware testing is inefficient and costly.
To address these challenges, we propose \sys~as a comprehensive solution.

\subsection{Preliminary Experiments}
\begin{figure}
    \centering
    \includegraphics[width=1\linewidth]{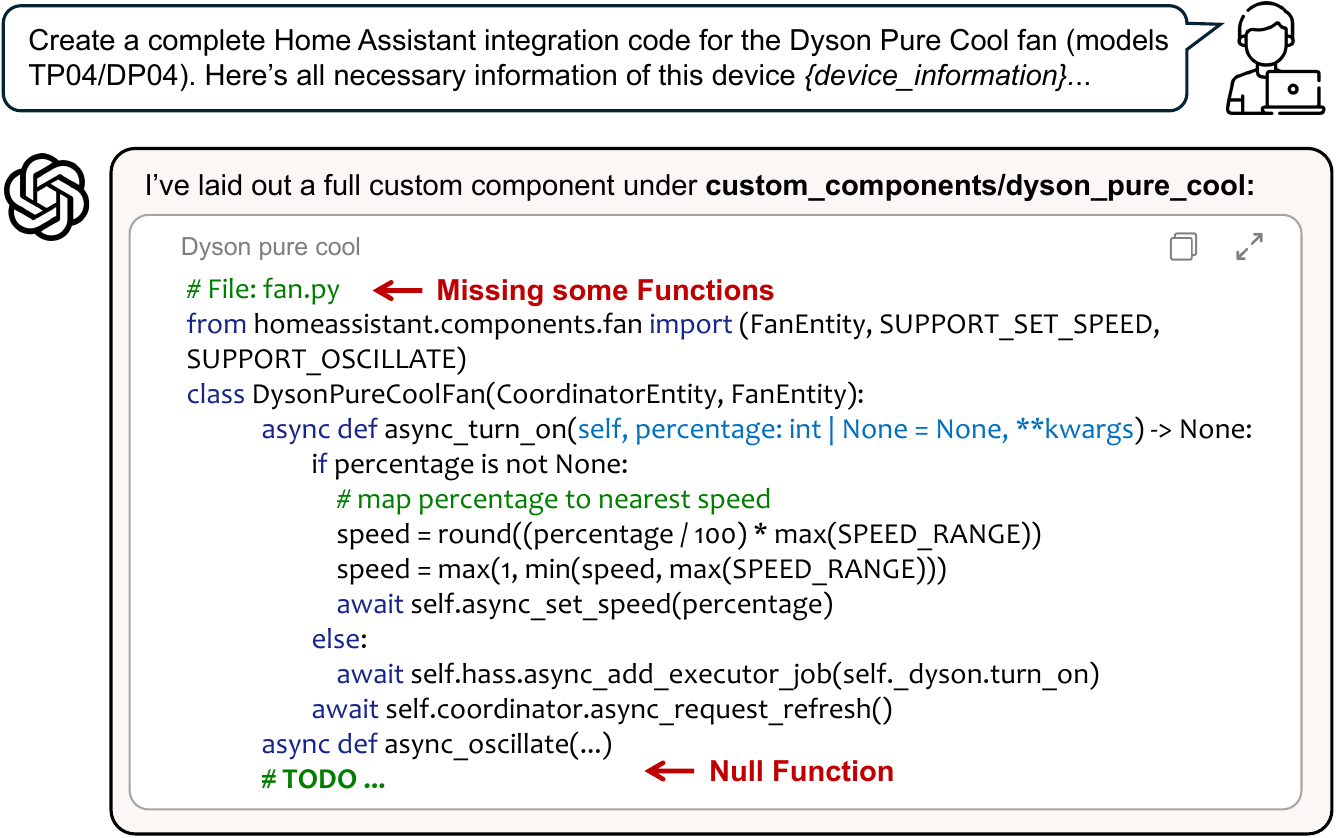}
    \caption{IoT Integration Generation with GPT-4o-mini-high}
    \label{fig:gpt-exp}
    \Description{}
    \vspace*{-6pt} 
\end{figure}

\begin{figure}
    \centering
    \includegraphics[width=1\linewidth]{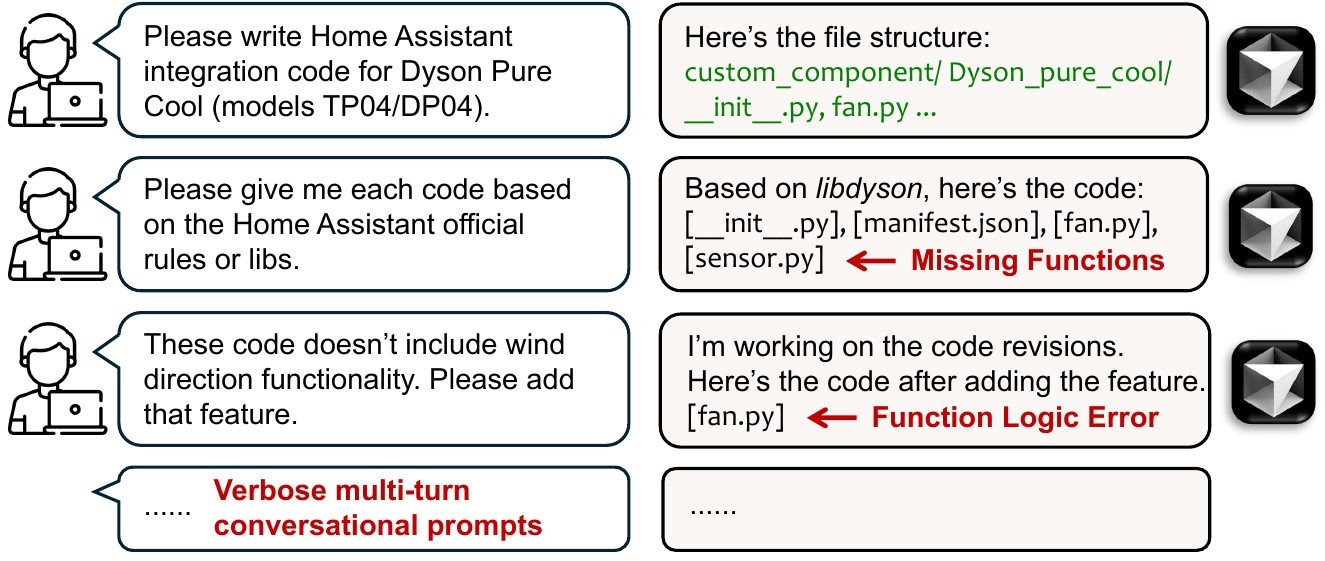}
    \caption{IoT Integration Generation with Coding Copilot}
    \label{fig:cursor-exp}
    \Description{}
    \vspace*{-12pt} 
\end{figure}

Before proposing our design, we first investigate an important question: \textit{Can state-of-the-art code LLMs—either standalone~\cite{claude4, gpto3, codellama-70b} or augmented with agentic architectures~\cite{cursor, copilot, codeassist}—effectively generate IoT integration code?}
Our preliminary experiments suggest a negative answer: advanced code LLMs and coding copilots fail to generate usable IoT integration code efficiently.
We use the integration code for Dyson fan as an example. 
A satisfactory integration code is expected to cover core functionalities of the fan, including power on/off, timer settings, speed control, oscillation, etc. 
As depicted in Figure~\ref{fig:gpt-exp}, when we instructed GPT-o4-mini-high~\cite{gpt-o4-mini-high} to generate the integration code, the result omitted several critical features—such as night mode, tilt control, and sleep timer, and even included placeholder functions marked with `\#TODO’, indicating incomplete implementation.
Similarly, we interacted with Cursor~\cite{cursor} to generate the code, as shown in Figure~\ref{fig:cursor-exp}.
The initial output was also functionally incomplete, and users must iteratively prompt Cursor through verbose multi-turn interactions to refine its responses. However, function logic errors and omissions persist in the generated code. Inexperienced developers rarely identify and address these issues, resulting in snowballing errors.

\begin{figure}
    \centering
    \includegraphics[width=1\linewidth]{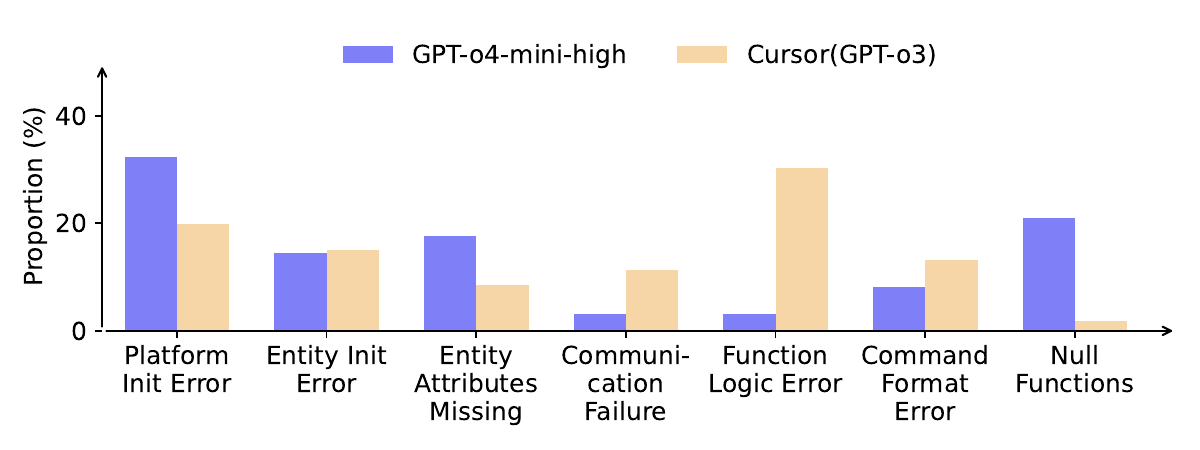}
    \caption{Error Types and Distribution}
    \Description{Error Types and Distribution in Preliminary Experiments.}
    \label{fig:gpt_cursor_comp}
    \vspace*{-14pt} 
\end{figure}

We tried the same task 10 times using GPT-o4-mini-high and Cursor, respectively, and examined their generated code to summarize the distribution of error types, as shown in Figure~\ref{fig:gpt_cursor_comp}.
Notably, the code generated by GPT exhibited numerous errors related to platform and entity specifications.
We hypothesize that this is due to the lack of specialized domain knowledge within the LLM—even the latest GPT models with web-based retrieval capabilities struggle to extract essential information from complex technical documentation.
Additionally, the high occurrence of null or placeholder functions indicates GPT’s inability to implement correct function logic. 
During our interaction with Cursor, we found that although persistent user prompting can incrementally improve the generated code, it is not trivial to fully eliminate all flaws automatically. In some cases, very specific hints from experienced programmers can guide the model toward a correct implementation. However, in other cases, it still fails to produce a correct result even after multiple iterations. 

In summary, although state-of-the-art LLMs offer long context windows, web-based retrieval capabilities, and convenient interactive interfaces for incorporating user feedback, generating workable IoT integration code efficiently remains challenging. It demands substantial domain knowledge and extensive interactive manual guidance. This motivates us to develop an automated solution that alleviates the development burden and facilitates the smooth expansion of multi-modal IoT serving systems, as shown in Figure~\ref{fig:intro-fig}.

\subsection{Motivation \& Key Ideas}

In this paper, we aim to develop an automated programming agent, \sys, to synthesize integration code for IoT devices.
Our key observation is that the complexity of integration code largely arises from its “bridge” nature.
Using the Dyson fan integration code as an example, as shown in Figure~\ref{fig:code-exp}, the code must correctly extend the abstract \texttt{FanEntity} defined by the platform, while also importing \texttt{libdyson} and invoking the appropriate device-specific functions.
To generate correct IoT integration code, the programming agent must possess both device-specific and platform-specific knowledge—and apply them accurately in the appropriate parts of the code. 
In addition, the “bridge” nature of IoT integration code implies that its correctness should ultimately be verified through actual hardware interaction—that is, by assessing whether the code enables the platform to fully control all functions of the IoT device. 

Based on the above observations, we design \sys~with three key modules: 
1)Integration Code Generator: equipped with divide-and-conquer reasoning instructions, it first generates device control logic and then refines it to be platform-compliant. During this process, it has access to knowledge retrieval tools and can autonomously decide whether, when, and what knowledge to retrieve to improve code generation accuracy. 
2) Automated Debugger: It registers a virtual device on the platform and composes test cases to validate whether the code complies with platform specifications. It can efficiently correct errors in an automated manner.
3) Hardware-in-the-Loop Debugger: It incorporates the real IoT device into the debugging loop to verify whether each function is correctly invoked. It requires only occasional binary feedback (“yes” or “no”) from the user based on the device’s physical behavior, while automatically handling error corrections.
\sys~provides a comprehensive solution designed to mitigate the practical development overhead associated with expanding multi-modal IoT systems—especially when serving a large number of customers with diverse preferences and budget considerations for device combinations in their own spaces. Ultimately, \sys~aims to make human-centered IoT services more flexible and affordable for more users. 


\section{System Design}
\label{section:Design}

\begin{figure*}[t]
    \centering
    \includegraphics[width=0.92\linewidth]{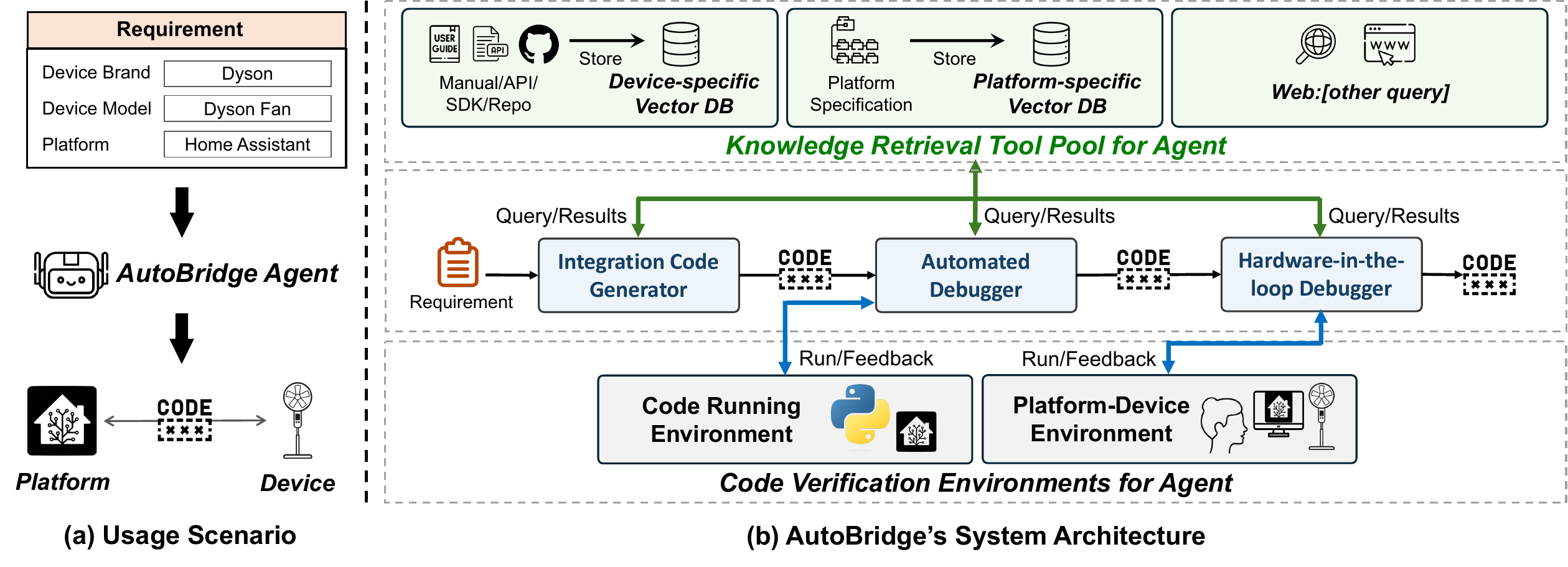}
    \caption{System Overview of \sys. 
    }
    \Description{}
    \label{fig:overview}
\end{figure*}

The system overview of \sys~is depicted in Figure~\ref{fig:overview}. 

\textbf{Usage Scenario.}
\sys~takes the user's specification of which device is to be integrated with which platform and generates the integration code that allows the platform to manage the device. 
As shown in Figure~\ref{fig:overview}(a), the user only needs to provide the \textit{device brand/model}, and \textit{platform}. 
For some specific devices, the user may also need to supply additional information, such as the device’s \textit{serial number}, \textit{key}, or a brief functional description.

\textbf{System Workflow.}
As shown in Figure~\ref{fig:overview}(b), \sys~consists of two main stages, code generation and code debugging, and features three modules. 
The first is the integration code generator, which retrieves both device-specific and platform-specific knowledge to assemble the integration code.
Based on the generated code, the second module, the automated debugger, registers a virtual device entity and debugs the code using a set of unit function tests within the code execution environment running the platform software. 
The third module, the hardware-in-the-loop debugger, further involves a real device and features an interactive approach to verify and debug the integration code until it can fully manage every function on the device, requiring only simple "yes/no" feedback from a human observer to indicate whether each function works correctly. The following sections elaborate on each individual module respectively. 

\subsection{Integration Code Generator}
\label{sec:Integration Code Generator}

\begin{figure}[t]
    \centering
    \includegraphics[width=0.95\linewidth]{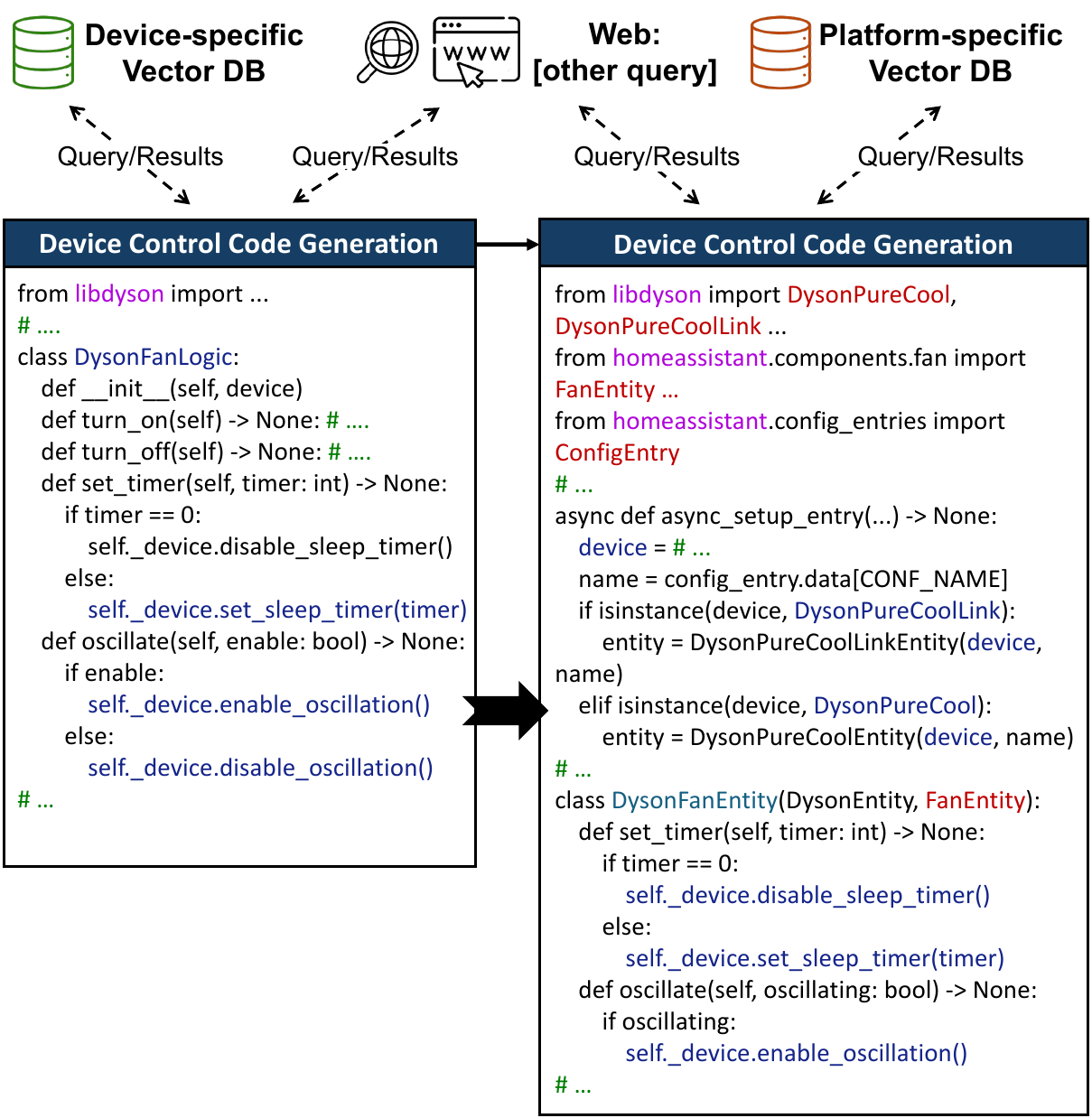}
    \caption{Workflow of Integration Code Generator. 
    It first retrieves device-specific knowledge to generate the \textit{Device Control Code}, and then further incorporates platform-specific knowledge to generate the complete \textit{IoT Integration Code}. 
    }
    \Description{}
    \label{fig:code-generator}
    \vspace*{-14pt} 
\end{figure}

\subsubsection{Task Overview}
Programming IoT integration code, even for expert programmers, is a domain-specific knowledge-intensive task.
Taking Figure~\ref{fig:code-exp} as an example, the code integrates a Dyson Fan into Home Assistant. The structure of this integration must follow the platform's specifications for adding a new device, including a \textit{manifest.json} file to specify metadata and one or more \textit{.py} files to implement the integration of each device function.
Within each \textit{.py} file, the code must incorporate the relevant abstract entities defined by the platform and correctly extend them, ensuring the integration is properly recognized by the platform software.
For instance, \texttt{DysonFanEntity} and \texttt{DysonNightModeSwitchEntity} are concrete implementations of the abstract entities \texttt{FanEntity} and \texttt{SwitchEntity} defined by Home Assistant.  
Importantly, the code should correctly incorporate the device's control interfaces and invoke them appropriately within the relevant functions of the device entities to implement actual control of the device. 
For instance, \texttt{enable\_night\_mode()} and \texttt{set\_sleep\_timer()} are concrete methods used to implement device control. 

The critical knowledge of how to control specific devices or what interfaces are available on target platforms is often missing or sparse in the intrinsic knowledge of code LLMs.
Prior work has demonstrated the effectiveness of RAG~\cite{gao2023retrieval, fan2024survey} in incorporating external knowledge to LLMs. 
However, blindly applying RAG does not fully address our challenges. First, the knowledge required for IoT integration code generation is twofold: both device-specific and platform-specific. When both knowledge are fed into the LLM simultaneously—even with a long context window capable of holding them—the resulting code generation accuracy remains unsatisfactory.
Second, different IoT devices have varying functional scopes, and a one-time fixed retrieval strategy often results in poor information efficiency and weak generalization across diverse IoT devices. 
We hypothesize that redundant information may increases the reasoning difficulty, making it harder for the LLM to apply the right knowledge in the right context.

\begin{figure*}[t]
    \centering
    \includegraphics[width=1\linewidth]{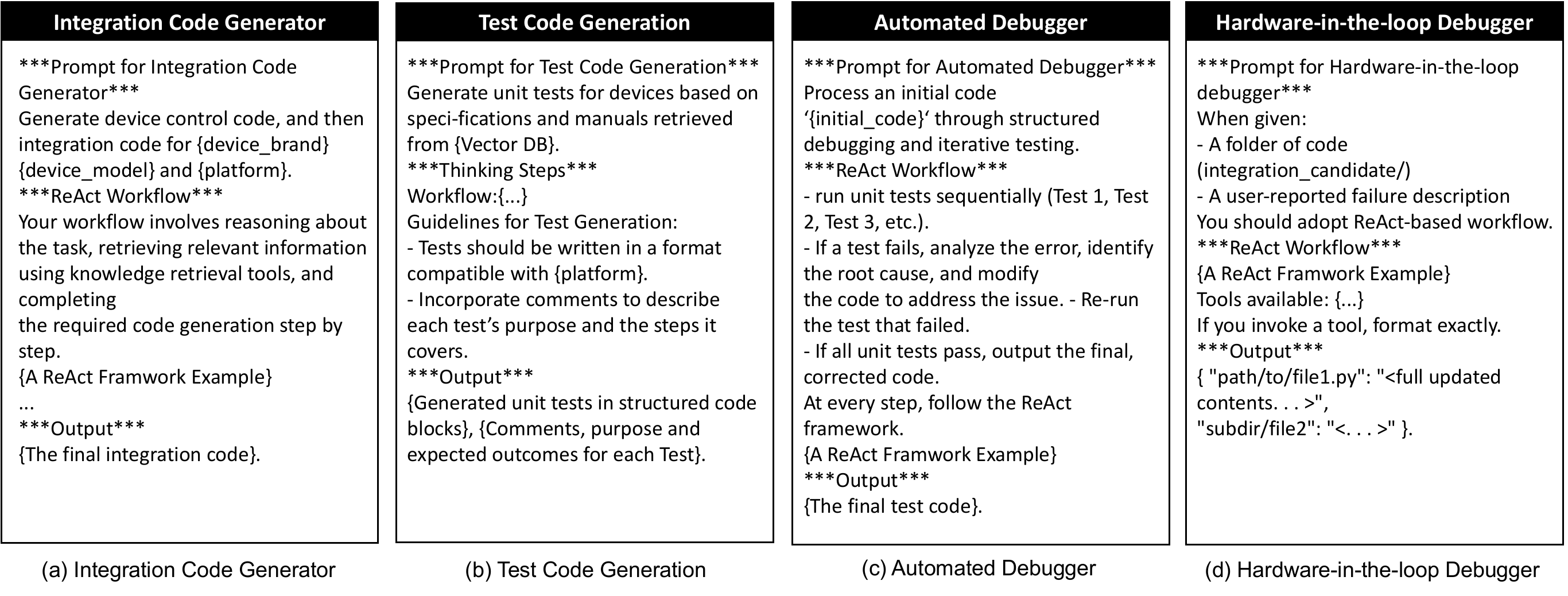}
    \caption{The prompt skeleton for (a) integration code Generator, (b) Test Code Generation, (c) Automated Debugger and (d) Hardware-in-the-loop Debugger.}
    \Description{}
    \label{fig:prompts}
\end{figure*}

\subsubsection{Progressive Knowledge Retrieval and Code Generation}
Due to the above issues, we divide the overall integration code generation into two sub-tasks, as illustrated in Figure~\ref{fig:code-generator}. First, it generates the device control code, focusing solely on device interfaces without considering adherence to platform specifications.
Then, based on the device control code, it further incorporates platform-specific knowledge to create a legitimate IoT integration on the targeted platform. 
The rationale behind this divide-and-conquer strategy is to prevent overwhelming the code generator with excessive knowledge from both sides at once, enabling it to generate more accurate code in the end. 
To flexibly handle the varying amounts of knowledge needed in program synthesis, we design a set of knowledge-retrival tools as follows:

\begin{itemize}[leftmargin=*]
    \item \textit{Device-Specific Vector DB}: Integration code needs to cover every function of the targeted device. To serve as a reference for the code generator, we use search APIs~\cite{googlesearchapi,githubsearchapi} to comprehensively retrieve the target device's user manuals, API/SDK documentation, and official GitHub repositories, which provide essential and reliable information about the device's functionality and relevant control interfaces. We then vectorize the retrieved data using embedding models~\cite{openaiembedding,bge_embedding}, and store them in vector database~\cite{douze2024faiss}. This is encapsulated as a tool, enabling the code generator and also debugger to query it as needed at any time.
    \item \textit{Platform-specific Vector DB}: Compared to device-specific knowledge, which spans multiple sources, platform-specific knowledge is mostly concentrated on the platform's official website. It typically includes a table of contents along with detailed descriptions of the platform's frameworks, entity class specifications, onboarding procedures for new integrations, and implementation guidelines~\cite{homeassistantExample}. Following the structure of the table of contents, we vectorize each leaf entry's content and store the resulting embeddings in a vector database~\cite{douze2024faiss}. This is also encapsulated as a tool for the code generator and the debugger.
    \item \textit{Web[Query]}: Besides the basic information, there is also other potentially useful knowledge scattered across the internet. To complement the Device/Platform-specific Vector DBs, especially for devices with more complex functionalities, we also encapsulate the web search API as a tool.
\end{itemize}

Given access to these knowledge-retrieval tools, we do not rigidly enforce the code generator to follow a fixed path for retrieving information. 
Instead, we guide the code generator using ReAct-based reasoning~\cite{yao2022react}. It interleaves reasoning and acting steps. 
Basically, we instruct the code generator to self-reflect on the generated code, and if it identifies missing knowledge, it should actively query for more information to assist the next steps.
At runtime, it has the autonomy to determine whether, what, and when to incorporate the necessary knowledge at each substep.
For example, during the device control code generation stage, it may query "the specific command format for turning on the device." Similarly, during the integration code generation stage, it may retrieve information such as "whether the Climate entity in Home Assistant contains the Set HVAC mode method." 
As shown in Table~\ref{tab:agenticRAG}, compared to a fixed, one-time retrieval approach, progressive retrieval driven by the underlying reasoning-acting loop can more effectively pinpoint informative knowledge for IoT integration code generation.

\begin{table}[t]
    \renewcommand{\arraystretch}{1.2}
    \small 
    \centering
    \begin{tabular}{|c|c|c|c|}
         \hline
         \textbf{Retrival Method} &  \textbf{\makecell[c]{Knowledge \\@Device Control \\ CodeGEN}} & \textbf{\makecell[c]{Knowledge \\@Integration \\ CodeGEN }} & \makecell{\textbf{Total}} \\
         \hline
         \makecell{Fixed\&One-time \\} & 1824 & 2197 & 4021 \\
         \hline
         \makecell{Progressive (ours)} & 1036 & 990 & 2026 \\ 
         \hline
    \end{tabular}
    \caption{Comparison between different knowledge retrieval strategies in code generation. (Unit: \#tokens) 
    }
    \label{tab:agenticRAG}
    \vspace*{-22pt} 
\end{table}


\begin{figure*}
    \centering
    \includegraphics[width=1\linewidth]{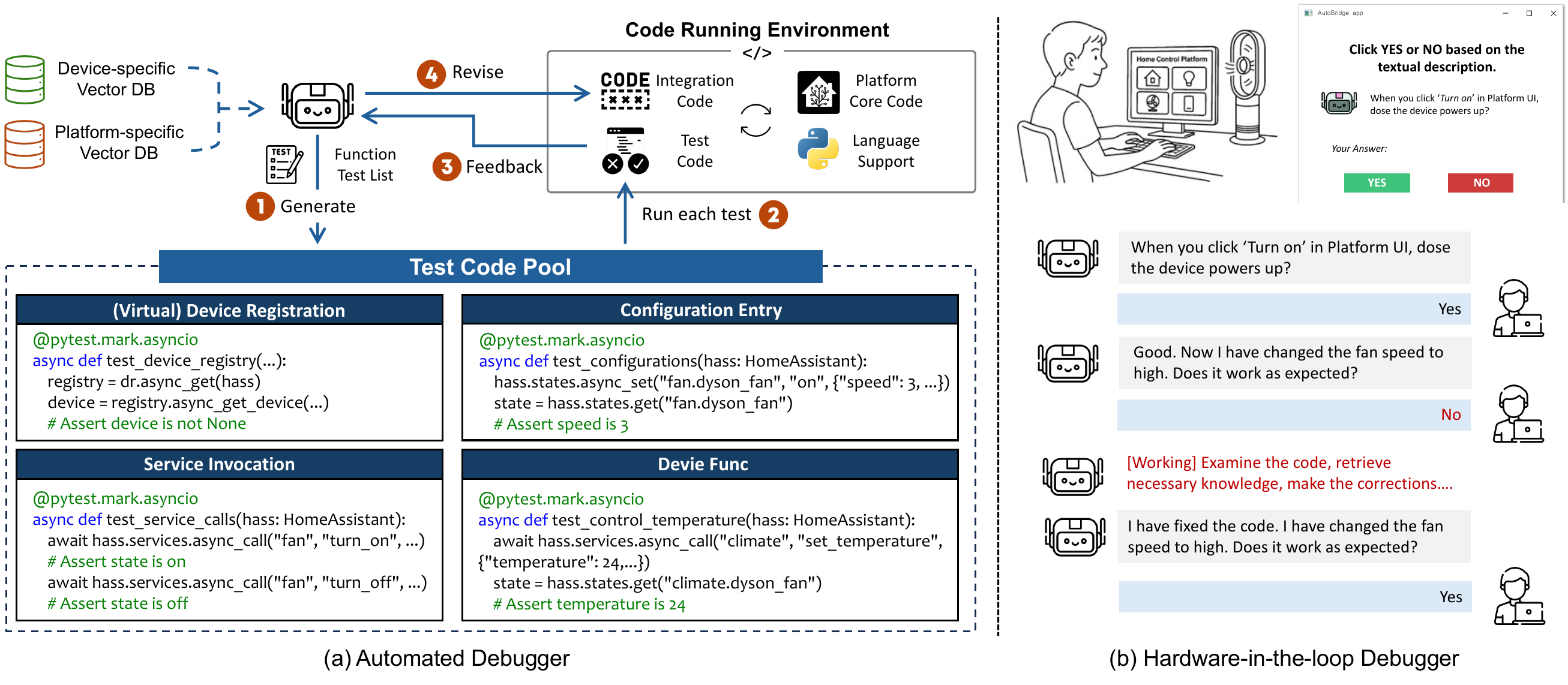}
    \caption{Workflow of AutoBridge's Multi-stage Debugging Framework. 
    }
    \label{fig:automated-debugger}
    \Description{}
    \vspace*{-6pt} 
\end{figure*}

\subsection{Automated Debugger}

\subsubsection{Task Overview}
Even with sufficient knowledge retrieval support, the generated code is not strictly guaranteed to be functional or to cover every function of the target device. Therefore, code verification and debugging must be systematically integrated into the workflow.

However, in our scenario, the challenge lies in the hardware-dependent nature of the IoT integration code. 
The gold standard for deeming integration code correct is that the deployed code must allow the platform to fully manage the device, meaning it can successfully invoke any function as desired.
Traditional static analysis tools can help us determine whether the generated code adheres to syntax and structural rules, but they cannot predict whether the code will truly allow the platform to fully manage the device and invoke functions as expected. This is because:
\begin{itemize}[leftmargin=*]
\item The interface between the device and platform requires strict adherence to subtle protocols and standards, which cannot be fully captured during the static analysis phases.
\item Issues such as timing problems, resource conflicts, or hardware-specific interactions can only be detected during dynamic validation, which occurs through debugging, testing, or manual inspection, not through static analysis.
\item The logic behind generating functional code may not align with user expectations. Inaccuracies in the code generator or unclear functional descriptions can lead to discrepancies between the generated code and the actual device functionality.
\end{itemize}

To address this, our solution is divided into two stages, ensuring code accuracy while minimizing costs from the user or developer. In the first stage, we use virtual/mock IoT devices to refine the code logic at the platform level. This stage is fully automated and does not require user intervention. In the second stage, we incorporate hardware-in-the-loop debugging to ensure that the code performs as expected on the actual device. The second stage will be explained in more detail in the next subsection (\S\ref{sec:Interactive Hardware-in-the-loop Debugger}).

\subsubsection{Virtual Device and Test Case Generation}
In the first stage, we design an automated debugger as illustrated in Figure~\ref{fig:automated-debugger}(a).
The basic idea is to have the generated code register a virtual device, which is then tested and refined within a coding environment that runs the platform's core software.
Despite not involving real hardware, this process allows us to verify whether the code is recognized as a legitimate integration from the platform's perspective, enabling us to fix relevant errors before moving on to the hardware testing stage. 
The virual device is an implementation of a full set of device interfaces, but its concrete implementations are enforced to return success signals or valid dummy values.

The comprehensiveness of the unit test code is crucial for the automated debugger.
In our design, we instruct the automated debugger to automatically create test cases. 
The pool of test code covers two main categories as follows:

\begin{itemize}[leftmargin=*]
    \item \textit{Basic Integration Test}: It mainly includes the following aspects: (a) (virtual) device registration: validate whether the device is successfully registered and appears in the platform registry; (b) service invocation: validate that functions, like \textit{turn on}, can be called; (c) configuration entry handling: simulate configuration entries to validate that the integration code can accept configuration changes. 
    \item \textit{Device Functionality Tests}: It validates that each device function is correctly bound to the corresponding platform service and can be invoked as expected. 
    \end{itemize}
    
We retrieve knowledge from the device- and platform-specific vector databases vector databases, summarize a device functionality list to serve as a reference for test code generation. 
In addition, we instruct the debugger to ensure the quality and clarity of the generated test code. 
Some generated code for testing integration between Dyson Fan and Home Assistant is shown in Figure~\ref{fig:automated-debugger}(a).
Given a pool of unit test code, we can now run them one by one in the code execution environment and revise the integration code based on the feedback from the test code execution. 
Here, we also guide the debugger with ReAct-based reasoning. 
The prompt skeletons of automated debugger are shown in Figure~\ref{fig:prompts}(b)(c).

It is noted that within the pool of test code, there may be few redundant or erroneous test, but this does not significantly impact the overall process.
Because we prioritize the comprehensiveness of the tests to cover as much device functionality as possible. This allows us to identify potential errors in the integration code early and correct them, ultimately reducing the time required for subsequent IoT hardware testing. 
Even if one individual test is erroneous, the ReAct-based reasoning may still correctly identify whether the issue stems from the test code itself rather than from the integration code.
Moreover, the following hardware-in-the-loop debugger serves as a fallback mechanism to ensure the final correctness of the IoT integration code. 

\subsection{Interactive Hardware-in-the-loop Debugger}
\label{sec:Interactive Hardware-in-the-loop Debugger}

To verify whether the integration code can ultimately interact correctly with real devices, we incorporate both the IoT device and a human observer into the debugging loop.

As illustrated in Figure~\ref{fig:automated-debugger}(b), the debugger has an interactive interface (graphical UI or terminal) that allows a human observer to provide feedback. 
The whole process is initiated and led by the debugger.
It first attempts to control a specific functionality, then prompts the human with a simple yes/no question, like "Did you observe the correct behavior?"
The human observer only needs to respond with "yes" or "no".
If the answer is "yes", the system proceeds to the next functionality.
If the answer is "no", the debugger will examine the code, possibly query the search tool for additional information, revise the code accordingly, and then resume testing from the same functionality. 
The debugger will continue the debugging process based on the feedback until every device function is verified to be working correctly. 

It is worth noting that our interactive interface can be easily extended to allow users to provide more detailed feedback.
However, in our current implementation, we stick to simple "yes/no" answers to ensure minimal user burden, enabling even non-programmers without technical expertise to interact smoothly with our system.
The prompt skeleton for the hardware-in-the-loop debugger is shown in Figure~\ref{fig:prompts}(d).

\section{Implementation}
\label{sec:implement}

We implement \sys~mainly with GPT-4~\cite{openai2023gpt}. 
We also evaluate the impact of different backbone LLMs using Gemini~\cite{gemini}, DeepSeek~\cite{deepseek-llm}, Claude~\cite{claude3}, and GPT-4o~\cite{gpt4o} in our ablation study.
For device-specific vector DB establishment, we use the Google Search Engine API~\cite{googlesearchapi} to retrieve the target device’s user manuals and API/SDK documentation, and the GitHub Search API~\cite{githubsearchapi} to retrieve the official repositories for the target device, retaining the top 5 items from each retrieval.
For textual content embedding, we use OpenAI’s \texttt{text-embedding-ada-002}~\cite{openaiembedding} model with its default settings (dimension 1536).
For code snippet embedding, we use BAAI’s\texttt{code-search-bge-base}~\cite{bge_embedding} model with default settings (dimension 768).
We adopt FAISS~\cite{douze2024faiss} with the IndexFlatL2 method~\cite{faissindex} for efficient similarity search of vector representations.
For platform-specific vector databases, we use the platform's developer community website as the knowledge source and apply the same embedding models and vector database tools as described above.
For two debuggers, the code verification environments accessed by them are deployed on a PC running Ubuntu 22.04 LTS and equipped with 2.10 GHz 13th Gen Intel Core i7-13700 CPU. 

\noindent\textbf{Disclaimer}. To ensure the rigor of our evaluation, we manually screen each piece of retrieved content from online sources, to exclude any exact integration code snippets for the device under evaluation.
This prevents accidental ground-truth leakage and ensures that our system generates integration code based on its comprehension and reasoning capabilities rather than copying and pasting existing solutions. 


\section{Evaluation}
\label{sec:benchmarkeval}

We aim to answer the following research questions:

\noindent \textbf{RQ1}: Can \sys~generate usable IoT integration code for a diverse range of IoT devices and platform combinations?

\noindent \textbf{RQ2}: Does the generated code enable the platform to control the full set of IoT device functionalities?

\noindent \textbf{RQ3}: How much human assistance is required to generate fully correct IoT integration code?

\noindent \textbf{RQ4}: How does each sub-design contribute to the overall performance of integration code generation?

\noindent \textbf{RQ5}: To what extent is the effectiveness of \sys~influenced by key factors such as backbone models and prompt granularity?


\begin{table*}[ht]
\renewcommand{\arraystretch}{1.1}
\small
\centering
\begin{tabular}{|c|c|c|c|}
\hline
\multicolumn{2}{|c|}{\diagbox{\textbf{EvalSet}}{\textbf{Platform}}}  & \textbf{Home Assistant} & \textbf{OpenHAB} \\
\hline
\multirow{3}{*}{\textbf{Real Hardware}} & Tier 1 & \multicolumn{2}{|c|}{Mi Smart Desk Lamp Lite, Mijia Smart Plug 3, Yeelight LED Bulb 1S (Color)} \\ 
\cline{2-4}
& Tier 2 &  \multicolumn{2}{|c|}{Aqara Smart Home Hub 2, Tuya Mini Multi-Mode Gateway} \\
\cline{2-4}
& Tier 3 &  \multicolumn{2}{|c|}{Ezviz C6c Camera, TP-Link DoorBell DB52C, Xiaoai Speaker Play Plus}  \\
\hline
\multirow{7}{*}{\textbf{HumanExpertCode}} & Tier 1 & Ikea Sensor & Tuya Switch \\ 
\cline{2-4}
& Tier 2 &  \makecell{ Asus Router, Netgear Router, KNX Floor Heating, \\ Rheem Water Heater, Anova Cooker, Dyson Fan, \\ Media Air Purifier, Petkit Feeding} & Sony Soundbar, EnOcean Gateway \\
\cline{2-4}
& Tier 3 &  \makecell{Epson Printer, TOSHIBA AC, Media AC, \\ Dahua Camera, Imou Camera, Dreame Vacuum, \\ Roomba Vacuum} & \makecell{ Eufy Doorbell, Media AC, \\ Hikvision Camera, Dahua Camera, \\ Mijia Robot Vacuum, Philips TV, Apple TV} \\
\hline
\end{tabular}
\caption{\sys's Evaluation Benchmarks. The devices with 1–6 functions are classified as Tier 1, those with 7–10 functions as Tier 2, and devices with more than 11 functions as Tier 3. 
}    
\label{tab:benchmarks}
\vspace*{-10pt} 
\end{table*}

\subsection{Evaluation Benchmark Construction}
Prior to our work, there was no readily available benchmark for IoT integration code generation tasks. Therefore, we constructed our benchmarks based on two popular open-source IoT automation platforms, Home Assistant~\cite{homeassistant2024} and openHAB~\cite{openHAB}. They are chosen because of their popularity within the open-source community. Moreover, Home Assistant is python-based, while openHAB is java-based. This diversity enables us to assess the generalizability of \sys~across different programming languages. Our benchmark covers two parts:

\begin{itemize}[leftmargin=*]
    \item \textbf{EvalSet 1: RealHardware}. We purchase 8 widely used IoT devices. With access to real hardware, we can directly deploy the generated integration code on the IoT devices for evaluation. This part allows both platforms to share the same evaluation set, facilitating fair comparison. 
    \item \textbf{EvalSet 2: HumanExpertCode}. Home Assistant and openHAB provide officially supported integrations developed by expert programmers. We select representative IoT integrations from each platform, 16 from Home Assistant and 10 from openHAB. In this category, each platform has its own dedicated evaluation set. This part complements the above RealHardware evaluation by incorporating a broader range of IoT device modalities to enrich the evaluation scope.
\end{itemize}

Devices in our benchmark vary in brand, modality, and functionality, as shown in Table~\ref{tab:benchmarks}.
Based on functional richness, we classify the smart devices into three tiers. 

\begin{itemize}[leftmargin=*]
    \item \textbf{Tier 1 Device (w/ simple functions)}: They often have few functions, mostly unary or binary, such as bulbs and switches. They generally require polling for control and information updates, as their functions are instantaneous rather than continuous.
    \item \textbf{Tier 2 Device}: These ones have more complex functions, covering gateways, routers, air purifiers and \etc.
    \item \textbf{Tier 3 Device (w/ complex functions)}: They are mostly those with multimedia functions, requiring continuous video streams, audio streams, \etc, for real-time monitoring and continuous commands, such as smart speakers, cameras, and cleaning robots.
\end{itemize}

Generally, IoT devices with more functionalities demand greater effort for programming its integration code. Therefore, our benchmark enables a comprehensive evaluation of \sys's capability to handle a wide range of complexities.


\subsection{Evaluation Metrics}

Following related efforts on code generation~\cite{dong2023self,shen2025autoiot}, we adopt the following evaluation metrics. 

\begin{itemize}[leftmargin=*]
    \item \textbf{Generation Success Rate (w/o Human Feedback)}. We run our method 30 times for each device-platform pair, then calculate the unbiased Pass@1 to assess the sucess rate of code generation. Formally, Pass@1 measures the probability that a single generated code sample (i.e., one attempt of the system) is an usable integration code. It quantifies \sys’s ability to generate usable code on the first try, without human feedback.
    \item \textbf{Functional Coverage (w/ or w/o Human Feedback)}. For usable code, this metric quantifies the percentage of functions correctly integrated relative to the full set of device functions.
    For the HumanExpertCode benchmark, we build function-specific test cases based on human-written code, test them on the generated code, and compare the responses between the generated and human-written versions for functional correctness verification.
    For the RealHardware benchmark, we directly test the generated code on the device and manually verify each individual function.
    \item \textbf{Number of Human's "no" Feedback}: This metric represents the number of "no" feedback responses from humans to assist the Hardware-in-the-loop Debugger. 
    It is used to evaluate the number of revision iterations assisted by humans. In general, fewer "no" feedback indicate a more effective integration generated by \sys’s fully automated components (the code generator and automated debugger). The maximum number of `no' feedback per device function is capped at 10.
\end{itemize}


\subsection{Performance w/o Human Feedback (RQ1/2)}

\begin{figure*}
    \centering
    \includegraphics[width=1\linewidth]{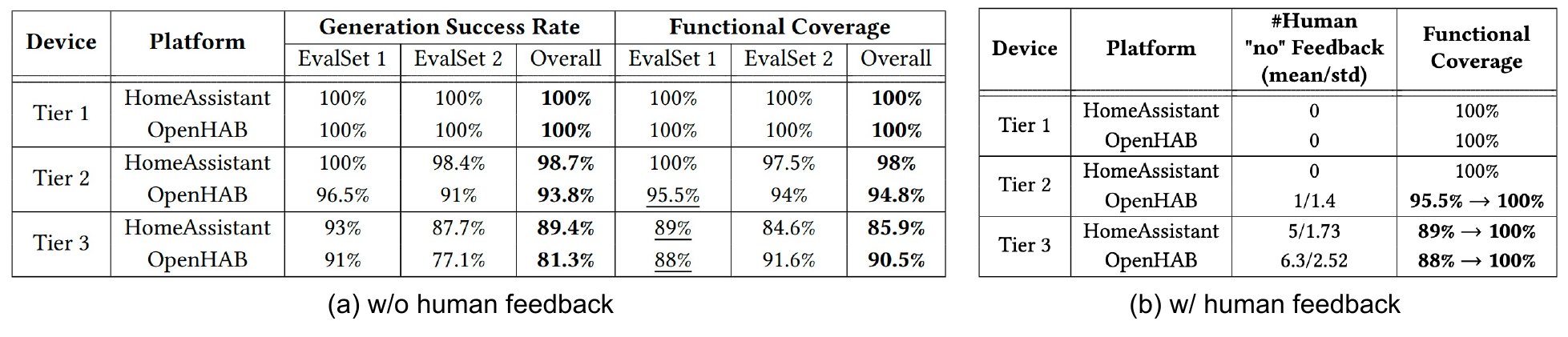}
    \caption{Benchmark-based Performance Evaluation: (a) \sys~(w/o Human Feedback)'s ability to synthesize integration code for various tiers of device complexity; (b) the amount of human feedback required to achieve fully correct integration code across different tiers of device complexity for EvalSet 1 (RealHardware).}
    \label{tab:eval_overall}
    \vspace*{-6pt} 
\end{figure*}

We first assess the code generation performance without human feedback, i.e., excluding the interactive hardware-in-the-loop debugger. In this evaluation, GPT-4 is used as the backbone LLM of \sys. The results across the whole benchmark are presented in Figure~\ref{tab:eval_overall}(a). 

\noindent\textbf{Performance@DeviceTier}
As the number of functions in a device increases, generating integration code generally becomes more challenging. In terms of generation success rate, the system experiences a bit performance drop when handling Tier 3 devices with more than 11 functions, but still maintains an average success rate above 80\%.
Functional coverage results are even more encouraging, among all usable code, on average, more than 85\% of device functions are correctly integrated.
These results demonstrate that \sys’s fully automated process (w/o human feedback) is effective to generate integration code for simpler devices, and covers most expected functionalities as device fucntion complexity increases.
It is also worth noting that the performance on EvalSet 1 is better than on EvalSet 2.
This is because we deliberately included more Tier 3 devices with complex functionality from HumanExpertCode to enrich the benchmark, making EvalSet 2 a harder code generation task. 

\noindent\textbf{Performance@Platform}
For Tier 1 devices, code generation performance on both platforms achieves 100\% in terms of generation success rate and functional coverage.
However, for most Tier 2 and Tier 3 devices, Home Assistant demonstrates a bit higher generation success rate than openHAB, 
The potential reason is that Home Assistant’s larger and more active user base contributes to a richer documentation ecosystem, providing more resources for \sys~to leverage during the code generation process. 
That said, a higher generation success rate does not necessarily imply higher functionality coverage. As shown in Figure~\ref{tab:eval_overall}(a), for Tier 3 devices, among all usable code, openHAB achieves 90.5\% functional coverage, while Home Assistant achieves 85.9\%.

\subsection{Performance w/ Human Feedback (RQ3)}

Based on the results in Figure~\ref{tab:eval_overall}(a), for the integration code without full functional coverage in EvalSet 1 (RealHardware), we further execute the interactive hardware-in-the-loop debugger.
Within the "no" feedback quota limitation, all integrations are ultimately able to achieve 100\% functional coverage, as shown in Figure~\ref{tab:eval_overall} (b). 
We also record the number of human "no" feedback instances for each device.
The value never exceeds 9 "no" responses, meaning no more than 9 rounds of code revision were needed to achieve correct integration code. 
More investigation regarding user experience will be illustrated in detail in the next section (\S\ref{sec:userstudy}).

\subsection{Ablation Study (RQ4/5)}
In this part, we conduct an ablation study to quantitatively evaluate the impact of different components in the \sys~design. This includes examining the impact of various backbone choices,  the impact of alternative generation prompts with varying levels of granularity, and the impact of sub-agent's design. This will provide insights into effectiveness of \sys~for future improvement.   

\begin{figure*}
    \centering
    \includegraphics[width=1\linewidth]{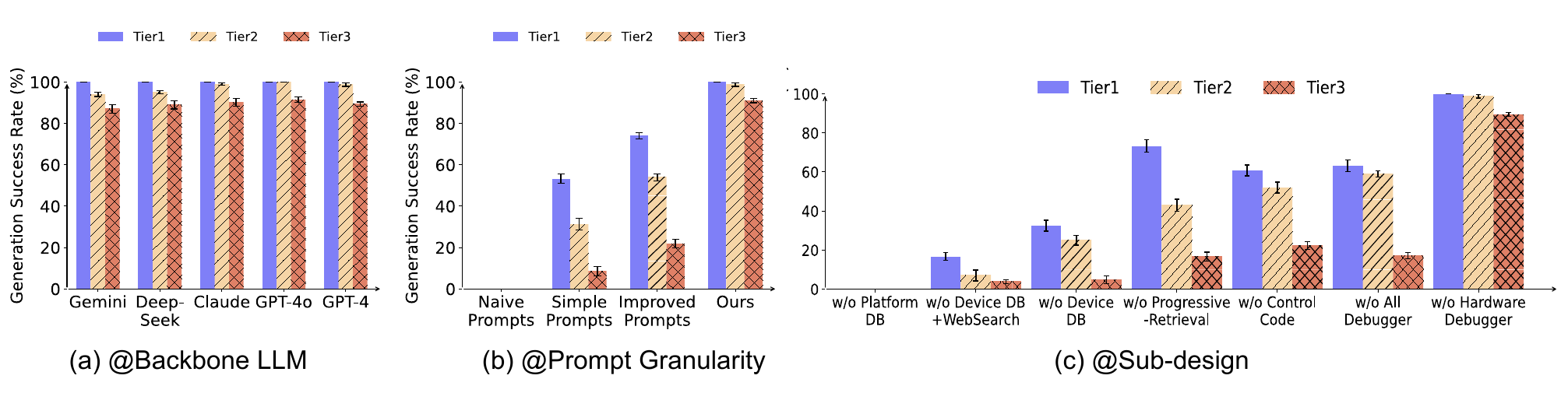}
    \caption{Ablation Study: (a) impact of different backbone LLM models; (b) impact of prompt granularity; (c) impact of sub-design on generation success rate of generated code on EvalSet 1 (RealHardware) benchmark.}
    \label{fig:ablation_combined}
    \vspace*{-6pt}
\end{figure*}

\noindent\textbf{Impact of Backbone LLM Model}
To investigate the impact of backbone models, we substituted the backbone of \sys's code generator and automated debugger with 4 state-of-the-art LLMs: Claude-3-Opus~\cite{claude3}, Gemini 1.5 Pro~\cite{gemini}, DeepSeek-V3~\cite{guo2024deepseek}, and GPT-4o~\cite{gpt4o}. 
Figure~\ref{fig:ablation_combined}(a) displays the generation success rate for various tiers of devices in EvalSet 1 (RealHardware) Benchmark. The results show that generation success rate for Gemini 1.5 Pro and DeepSeek-V3 were lower than those of GPT-4 to varying degrees, while generation success rate using the more recently released Claude-3-Opus and GPT-4o as the backbone was slightly higher than those obtained using GPT-4. Overall, replacing the backbone did not affect the generation success rate of generated code by more than 4.5\%. This indicates that the performance of \sys~is relatively stable regardless of the underlying LLM backbone used.

\noindent\textbf{Impact of Prompt Granularity}
Here we explore the impact of the granularity of prompts used for the code generator.
We compare four types of prompts for the code generator:
(1) a naive prompt with only a basic task description;
(2) a simple prompt that mentions the availability of knowledge retrieval tools;
(3) an improved prompt with additional explanations on how to use the knowledge retrieval tools; 
(4) A full-version prompt (ours) with detailed instructions on the ReAct-based reasoning strategy.
Figure~\ref{fig:ablation_combined}(b) shows that the granularity of the prompts significantly affects the generation performance. The naive prompt is entirely unable to generate working code. As the granularity of the prompts increases, the generation success rate of generated code improves significantly across different tiers of devices. For the first, second, and third tier devices, the improvements in generation success rate from simple prompts to full-version prompts are 46.7\%, 68.7\%, and 91.5\% respectively.
This highlights the critical role that well-crafted prompts play in optimizing code generation systems like \sys. By carefully designing prompts to include more detailed and specific guidance, the system can better understand and execute the complex tasks required for effective integration generation. 

\begin{figure*}[t]
    \centering
    \includegraphics[width=0.92\linewidth]{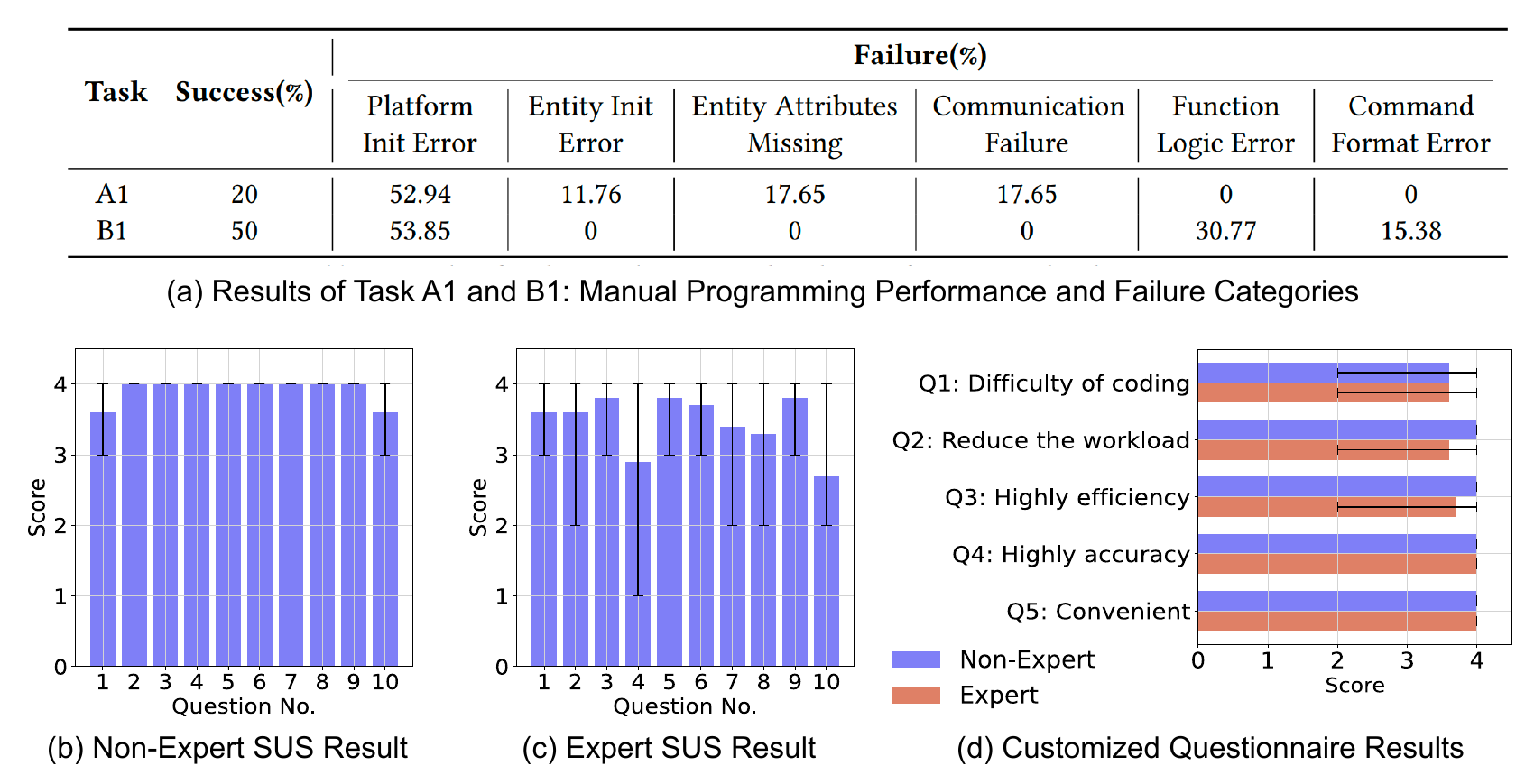}
    \caption{\sys's User Study}
    \label{fig:userstudy}
\end{figure*}

\noindent\textbf{Impact of Sub-Design}
To investigate the importance of the sub-agents' design for the code generation performance, we evaluate and compare the following versions: (1) w/o platform-specfic vector DB; (2) w/o device-specific vector DB and web search; (3) w/o device-specific vector DB; (4) w/o progressive knowledge retrival capability; (5) w/o intermediate device control code generation; (6) w/o automated and hardware-in-the-loop debugger; (7) w/o hardware-in-the-loop debugger. 

(a) \textbf{Impact of Accessible Knowledge}.
As shown in Figure~\ref{fig:ablation_combined}(c), excluding the platform-specific DB completely prevents the system from producing usable results, while excluding the device-specific DB leads to a decrease in generation success rate ranging from 67.5\% to 86.2\%.
This implies that platform-specific knowledge is likely more domain-specific and less present in the intrinsic knowledge of LLMs.
The absence of progressive retrieval capability has a particularly significant impact on Tier 2 and Tier 3 devices with more functions, resulting in generation success rate reductions of 55.7\% and 74.3\%, respectively.
These results highlight the knowledge-intensive nature of integration code generation and the necessity of the knowledge retrieval tools design in \sys. 

(b) \textbf{Impact of Intermediate Device Control Code}.
As shown in Figure~\ref{fig:ablation_combined}(c), skipping the step of generating device control code negatively impacts the generation success rate, with performance drops exceeding 66.7\% for Tier 3 devices.
This implies that if the code generator does not adopt the divide-and-conquer strategy, attempting to retrieve both device- and platform-specific knowledge and generate the integration code in one step, the resulting outputs are unsatisfactory, especially for devices with complex functions.
This validates the effectiveness of our code generator design. 

(c) \textbf{Impact of Debugger}.
As shown in Figure~\ref{fig:ablation_combined}(c), removing both the automated debugger and hardware-in-the-loop debugger negatively impacts the generation success rate.
Even without hardware involvement, preserving the automated debugger can still largely maintain a high generation success rate, especially for Tier 1 and Tier 2 devices.
This validates the effectiveness of our automated debugger design, showing that the automated debugger can greatly reduce the burden on human involvement during the subsequent hardware-in-the-loop debugging process. 
\section{User Study}
\label{sec:userstudy}

In addition to benchmark-based performance evaluation, we conduct a user study to assess the usability of \sys.
We develop a graphical UI application that enables users to interact with \sys.
We evaluate the efficiency of \sys~compared to expert-driven coding and, more importantly, assess its usability through user feedback to answer the following questions: 

\noindent \textbf{RQ6}: What are the perceived difficulty of programming IoT integration code for people with varying programming expertise?

\noindent \textbf{RQ7}: Compared to manual coding, how accurate and efficient is \sys~in generating IoT integration code?

\noindent \textbf{RQ8}: Does \sys~provide a seamless user experience, regardless of their backgrounds and programming expertise?

\subsection{Study Procedure}

\textbf{Participants}.
We recruited 15 participants (ages M = 23.6 years, SD = 3.4 years): (1) \textit{Non-Expert Group}: 5 participants w/o educational background in EE/CS and programming expertise. (2) \textit{Expert Group}: 10 participants with EE/CS background and programming expertise, and also with domain knowledge about embedded systems development ranging from 0 to over 3 years. 


\noindent\textbf{Evaluation Task}.
We selected two pairs of platform and device to construct the integration tasks: (1) \textit{Task A}: Integrating a self-developed smart thermo-hygrometer device with the Home Assistant platform. (2) \textit{Task B}: Integrating a Yeelight LED Bulb 1S (Color)~\cite{yeelightled} with the Home Assistant platform.
Device in Task A is a self-developed device composed of an ESP8266 Wi-Fi module, a 0.96-inch OLED display, and a DHT11 sensor. It has two main functions: (a) \textit{Update}: updating real-time temperature and humidity readings on the display, and (b) \textit{Information Transmission}: communicating data via HTTP, mimicking the functionality of a commercial product, Xiaomi Mijia Bluetooth Thermo-Hygrometer~\cite{xiaomimonitor}.
Despite its functional simplicity, the integration code must still implement platform/entity initialization, service calls, and configuration entries, making it a suitable task for the user study. 
Device in Task B is a commercial product with more functionality—such as turn on, turn off, brightness control, color temperature control, color switching, and rhythmic mode. It also has more detailed documentation than Task A, ensuring task diversity. 


\noindent\textbf{Study Process}.
Each participant was provided with a computer pre-configured with Home Assistant and IoT devices for Task A/B.
Participants were instructed to complete the following steps:
\begin{enumerate}[leftmargin=*]
    \item \textbf{Demographic Questionnaire}: Collect participants' demographic information, particularly their background expertise.
    \item \textbf{Background Learning} (no time limits, \textasciitilde20 minutes): Each participant followed a detailed tutorial to learn background knowledge and task details.
    \item \textbf{Task A1: Manually Complete Task A} (expert group only, max 40 minutes): Participants wrote the desired IoT integration code for Task A.
    They can use any internet-based LLMs.
    \item \textbf{Task A2: Using AutoBridge to Complete Task A} (no time limits, \textasciitilde25-30 minutes): Participants used AutoBridge to generate the desired IoT integration code. 
    \item \textbf{Task B1: Manually Complete Task B} (expert group only, max 40 minutes). 
    \item \textbf{Task B2: Using AutoBridge to Complete Task B} (no time limits, \textasciitilde25-30 minutes). 
    \item \textbf{Post-Task Questionnaire} (5 minutes)
    \item \textbf{Semi-Structured Interview} (10–15 minutes)
\end{enumerate}

Between Task A and Task B, participants were allowed to take a break. 
After the study, participants were compensated at a rate of \$12 per hour.
This study has been approved by the Institutional Review Board
(IRB) of our institution. 



\subsection{User Study Results}

\textbf{Manual Programming Performance (RQ6)}.
We first assess the manual coding performance of the expert group to investigate the perceived difficulty of programming integration code.
As shown in Figure~\ref{fig:userstudy}(a), the average success rates for Task A1 and Task B1 are 20\% and 50\%, respectively, indicating that it is not an trivial task, even for programmers working with simple IoT devices. 
We also examined the unsuccessful code submissions and categorized their failure types.
As shown in Figure~\ref{fig:userstudy}(a), more than 50\% of the failures were attributed to platform initialization errors for both tasks.
Many participants, even those with embedded programming expertise, struggled to correctly implement code adhering to platform specifications.
In Task A1, although the self-developed device was functionally simple, it lacked mature documentation, resulting in more errors related to entity initialization, entity attribute configuration, and communication handling.
In contrast, Task B1 involved a commercial device with comprehensive API documentation. However, due to the device’s richer functionalities, common failures included function logic errors and incorrect command formatting. 

\noindent\textbf{\sys~ v.s. Manual Programming (RQ7)}.
For Task A2 and B2, \sys~successfully generated the correct integration codes within 30 minutes. 
This implies that in terms of code accuracy, our system outperforms programmers’ performance by 50\%–80\%, even when programmers are allowed to use commercial code LLMs. 
For the self-developed device, we recorded the internal reasoning process of \sys~and found that its ReAct-based strategy guided it to first analyze the input function description of the device. It then reasoned that such functionality is likely implemented through HTTP requests with a specific JSON format, invoked the search tool to retrieve relevant knowledge, and ultimately generated the correct integration code.

\noindent\textbf{Subjective Measurements (RQ8)}.
We employed System Usability Scale (SUS)~\cite{bangor2008empirical} questionnaire, which is widely used to evaluate system usability.
The standard SUS questionnaire consists of 10 items rated on a 5-point Likert scale.
We computed the raw scores for each participant following the scoring method described in~\cite{DBLP:journals/corr/abs-2007-15951}, i.e., converting the range from 1–5 to 0–4.
The results are shown in Figure~\ref{fig:userstudy}(b) and(c).
The non-expert group achieved an SU score of 98.00 (± 3.00), with a usability subscore of 98.75 (± 1.875).
The expert group reported an SU score of 86.50 (± 11.50) and a usability subscore of 90.625 (± 9.375).
These results indicate that \sys~is highly usable, demonstrating ease of task completion across users with varying technical backgrounds. 
In addition to the standard SUS questionnaire, 
we also designed an additional questionnaire with the following questions: (1) The manual coding task was very difficult; (2) \sys~significantly reduced my workload compared to manual coding; (3) \sys~was highly efficient; (4) \sys~demonstrated a very high level of accuracy; (5) \sys~was very easy to use and not cumbersome.
all questions were answered using a 5-point scale to indicate participants' level of agreement with the statements, where the lowest score means "strongly disagree" and the highest score means "strongly agree".
The results are shown in Figure~\ref{fig:userstudy}(d).
Most participants reported that manually writing integration code was a highly challenging task and that \sys~significantly reduced their workload.

\section{Discussion}
\label{sec:discuss}


\noindent\textbf{Enhancing Developer Productivity and Learning}. 
In \S\ref{sec:userstudy}, we demonstrated \sys’s ability to boost human developers’ efficiency and accuracy. 
But some developers may require more than a “black-box” solution that simply outputs final code. 
They seek visibility into each step of the reasoning process to support self-directed learning and to audit decision rationale. 
Future work could provide a clear, interactive, step-by-step interface~\cite{waitgpt} that exposes the LLM’s decision logic, thereby serving as a general-purpose educational tool for developers.

\noindent\textbf{Streamlining Hardware-in-the-Loop Debugger}.
The hardware-in-the-Loop (HIL) scheme was introduced to provide a high-fidelity approach for debugging hardware-dependent code, such as IoT integration code. 
In \sys, we minimize the effort required for the HIL debugger, only providing simple binary feedback.
Recent advances in vision–language models~\cite{xu2021vlm}, such as Qwen2.5-VL~\cite{Qwen2.5-VL} and DeepSeek-VL2~\cite{wu2024deepseekvl2mixtureofexpertsvisionlanguagemodels}, are promising for understanding device status through visual inputs, making HIL even more efficient and potentially human-free.


\noindent\textbf{Generalization to Knowledge-Intensive Hardware-Dependent Code Generation}.
It is noted that many domain-specific hardware-relevant code generation tasks share a knowledge-intensive nature and rigorous hardware testing requirements. 
Based on \sys, future research could explore a extensible framework, standardizing domain-knowledge retrieval protocols and abstracting hardware testing interfaces to make adaptation to other domains easier. 


\section{Conclusion}
\label{sec:conclude}

In this paper, we present \sys, an LLM-powered code agent designed to automate the synthesis of IoT integration code.
Our design is driven by the deep investigation into the nature of IoT integration code. We design a divide-and-conquer strategy that enables the code agent to incorporate domain knowledge effectively during different stages of code generation. In addition, we design a multi-stage debugging framework to ensure the correctness of the generated code. Our comprehensive evaluation demonstrates the effectiveness, efficiency, and usability of \sys.
\sys\ significantly reduces development overhead when incorporating new IoT devices into multi-modal IoT system, accelerating the process of meeting customers’ diverse preferences in device combinations to deliver human-centered services.


\bibliographystyle{ACM-Reference-Format}
\bibliography{mybib}

%

\end{document}